\documentclass{ws-ijmpa}

\begin{document}

\markboth{GENG AND HSIAO}
{T VIOLATION IN BARYONIC B DECAYS}

%
\catchline{}{}{}{}{}
%
\def\ra{\rightarrow}
\def\be{\begin{equation}}
\def\ee{\end{equation}}
\def\bea{\begin{eqnarray}}
\def\eea{\end{eqnarray}}
\def\lln{\left<}
\def\rln{\right>}
\def\gam{\gamma}
\def\prl{Phys. Rev. Lett.~}
\def\pr{Phys. Rev.~}
\def\pl{Phys. Lett.~}
\def\np{Nucl. Phys.~}
\def\prl{Phys. Rev. Lett.~}
\def\tp{\vec v_1 \cdot (\vec v_2 \times \vec v_3)}

\title{T VIOLATION IN BARYONIC B DECAYS
}

\author{\footnotesize C.~Q.~GENG
 AND Y.~K.~HSIAO}
\address{Department of Physics, National Tsing Hua University,\\
Hsinchu, Taiwan 300}

\maketitle


\begin{abstract}
We study $T$ violation in the three-body charmless baryonic decay
of $\bar B^0 \ra \Lambda \bar p \pi^+$ through the $T$-odd triple
product correlation in the standard model.  We find that the $T$ violating asymmetry is
$10\%$, which is accessible to the current B factories at KEK
and SLAC, while the $CP$ violating one $1.1\%$.
We emphasize that this triple product correlation
would be the first measurable direct
$T$ violating effect predicted in the standard model.

\keywords{CP and T violation; B meson decays; Baryonic modes.}
\end{abstract}

\vspace{1.cm}
 CP violation induced by the $B-\bar{B}$ mixing
in the CKM framework of the standard
model, namely $\sin 2\beta$, has already been measured by Belle
and Babar collaborations at KEK and SLAC, respectively
\cite{belbab,Acp}. 
Other characteristic observables of CP violation are rate asymmetries
and momentum correlations. The direct
CP asymmetries
 arise if both the weak ($\gamma$) and strong
($\delta$) phases are non-vanishing 
\be A_{CP}  \propto \sin
\gamma \sin \delta. \label{cpa} 
\ee
 Whereas the correlations among
spin and momenta of the intial and final state  particles
constitute a measure of $T$-violating observables.
 The correlations known as triple product
correlations (TPC's), of the $T$-odd form $\vec v_1\cdot(\vec v_2
\times \vec v_3)$, where $\vec v_i$'s are spin ($\vec{s}_i$) or momentum ($\vec{p}_i$), are
used to probe $T$-violation.
In the framework
  of local quantum field theories, 
  T-violation implies CP-violation (and vice versa), because of the
  CPT invariance of such theories. 
  Experimentally, T violation has been only observed in the neutral kaon system \cite{TV} so far.
Moreover, 
  no violation of CPT symmetry has been found~\cite{pdg}.
  Still, it will be worthwhile to
  remember that outside this framework of local quantum field
  theories, there is no
  reason for the two symmetries to be linked~\cite{CPT}. 
  Therefore, it would be
  interesting to directly investigate T violation in B decays,  rather than infering
  it as a consequence of CP-violation.

A nonzero TPC is given by 
\be A_T = {\Gamma (\tp > 0)
- \Gamma (\tp < 0) \over {\Gamma (\tp
> 0) + \Gamma (\tp < 0)}}\;,
\label{atp}
\ee
where $\Gamma$ is the
decay rate of the process in question. In comparison with the
conjucate process, TPC asymmetry (TPA), ${\cal A}_T$ is expressed
as \be {\cal A}_T = {1 \over 2}(A_T-\bar A_T). \label{tpa} \ee By
expressing so, we reaffirm the TPC is indeed due to weak phase.
Otherwise, the nonzero TPC in Eq. (\ref{atp}) can occur due to
only strong phase. Then TPA turns out to be:
\be
 {\cal A}_T
\propto \sin \gamma \cos \delta.
 \ee
  This is in contrast with the CP
asymmetry in Eq. (\ref{cpa}).  In the
vanishing limit of the strong phase, the TPA is maximal.
 We note that there is no contribution to
${\cal A}_T$ in Eq. (\ref{tpa}) from final state interaction due
to electromagnetic interaction.
In this talk, we consider the three-body charmless baryonic
process \cite{GHT1} of $\bar B^0 \ra \Lambda \bar p \pi^+$ looking for TPC of
the type $\vec s_{\Lambda} \cdot (\vec p_{\bar p}\times\vec
p_{\Lambda})$. 
Our objectives are to test the CKM
paradigm of CP violation and unfold the
physics beyond it.

It is interesting to note that the branching
ratio of three-body baryonic decay is much larger than that of the
two-body one with the same baryon pair as observed~\cite{belle0,belle2}: 
\bea\label{exbr}
 Br( B^0 \ra \bar{\Lambda} p\pi^-) =
(3.27^{+0.62}_{-0.51}\pm0.39) \times 10^{-6}\,,\ 
Br(B^- \ra\Lambda \bar p) &<& 4.6 \times 10^{-7}\,.
\eea The enhancement of three-body decay over the two-body one is
due to the reduced energy release in $B$ to $\pi$ transion by the
fastly recoiling $\pi$ meson that favors the dibaryon production
\cite{soni}. Theoretical estimations on the modes in Eq.
(\ref{exbr}) and other baryonic B decays are made
\cite{chua,group,GH05}, in consistent with
the experimental observations.

In the factorization method, the decay amplitude of $\bar B^0 \to
\Lambda \bar p \pi^+$ contains the $\bar B^0\to\pi^+$ transition
and $\Lambda \bar p$ baryon-pair inducing from the vacuum. The
contributions to the decay at the quark level are mainly from
$O_1$, $O_4$ and $O_6$ operators defined in Ref.
\cite{ali}. From those operators and the factorization
approximation, the decay amplitude is given by \cite{GHT1,chua}
\begin{eqnarray}\label{M146}
M&=&M_1+M_4+M_6\nonumber\;,\\
M_i&=& {G_f \over \sqrt{2}} \lambda_i a_i \lln \pi^+| \bar u
\gamma^\mu(1-\gamma_5)b|\bar B^0\rln \lln \Lambda \bar p|\bar s
\gamma_\mu(1-\gamma_5)u|0\rln
\label{m1}\;,\ (i=1,4)\,,\nonumber\\
M_6 &=& {G_f \over \sqrt{2}} V_{tb} V_{ts}^* 2a_6 \lln \pi^+ |
\bar u \gamma^\mu(1-\gamma_5)b|\bar B^0 \rln {(p_\Lambda +p_{\bar
p})_\mu \over {m_b - m_u}} \lln \Lambda \bar p|\bar s
(1+\gamma_5)u|0\rln \label{m6}\;,
\end{eqnarray}
where $\lambda_1=V_{ub} V_{us}^*$ and $\lambda_4=-V_{tb} V_{ts}^*$
and
\begin{eqnarray}\label{a146}
a_1=c_1^{eff}+\frac{1}{N_c}c_2^{eff}\;,\;a_4=c_4^{eff}+\frac{1}{N_c}c_3^{eff}\;,\;a_6=c_6^{eff}+\frac{1}{N_c}c_5^{eff}\;,
\end{eqnarray}
with
$c_i^{eff}\;(i=1,2, ..., 6)$ being effective Wilson coefficients
(WC's) given in Ref.~\cite{ali} and
$N_c$ color number.
In Eq. (\ref{a146}), we note that $c_i^{eff}/N_c$ are included to
express the color-octet terms. 
The hadronic matrix elements in Eq. (\ref{a146}) can be found in Ref.~\cite{GHT1}.

We now study the TPC involving the $\Lambda$ spin from Eq.
(\ref{M146}), and we get
\begin{eqnarray}
|M|^2 = |M_1|^2 + |M_4|^2 + |M_6|^2 + 2 Re (M_1 M_4^\dagger) + 2
Re (M_1 M_6^\dagger) + 2 Re (M_4 M_6^\dagger)\;.
\end{eqnarray}
We note that
TPC can only arise from the interference terms, ${\it i.e.}$,
$Re(M_1 M_4^\dagger)$, $Re(M_1 M_6^\dagger)$ and $Re(M_4
M_6^\dagger)$. However,  the $T$-odd term in
$Re(M_1 M_4^\dagger)$ disappears since $M_1$ and $M_4$ have the
same current structures as seen from Eq. (\ref{m6}).
 Explicitly, the T-odd transverse polarization asymmetry
$P_T$, which is related to TPC, is found to be
\begin{eqnarray}
P_T=\frac{8G_F^2\;m_B|\vec{p}_{\bar p} \times
\vec{p}_\Lambda||V^{\;}_{tb}V^{*}_{ts}|^2}
{\rho_0} \bigg[\bigg(V\cdot S+A\cdot
P\bigg)Im\left({V^{\;}_{ub}V^{*}_{us}\over
V^{\;}_{tb}V^{*}_{ts}}a_1^{\;}a_6^*
               -a_4^{\;}a_6^*\right)\bigg]\,,
\end{eqnarray}
where $S,P,V$ and $A$ are combinations of form factors, defined in Ref.~\cite{GHT1}.
It is noted that the $V\cdot S$ ($A\cdot P$) term is from
vector-scalar (axialvector-pseudoscalar) interference and there is
no T-odd term from $Re(M_1 M_4^\dagger)$ due to the same current
structures. We may also define the integrated
transverse $\Lambda$ polarization asymmetry $A_T$ in terms of Eq.~(\ref{atp}).

In our numerical calculations, the CKM parameters are taken to be
\cite{pdg} $V^{\;}_{ub}V^{*}_{us}=A\lambda^4(\rho-i\eta)$ and
$V^{\;}_{tb}V^{*}_{ts}=-A\lambda^2$ with $A=0.853$,
$\lambda=0.2200$, and $\rho$ and $\eta$ are expressed as functions
of the weak phase $\gamma$ by $\rho=R_b\;\cos\gamma\;$ and
$\eta=R_b\;\sin\gamma$ with
$R_b\equiv\frac{|V_{ub}|}{|V_{cb}|}/\lambda=0.403$ \cite{pdg}. We
note that the current allowed values of  $(\rho,\eta)$ are
$(0.20\pm0.09,0.33\pm0.05)$ \cite{pdg}. To distinguish the origin
of CP violation, we use the weak phases $\gamma=60^\circ$ and
$0^\circ$ corresponding to $(\rho,\eta)=(0.20,0.35)$ and
$(0.40,0)$, respectively. We remark that $a_1$, $a_4$ and $a_6$
\cite{ali} contain both weak and strong phases,
induced by $\eta$ and quark-loop rescatterings, respectively.
Explicitly, at the scale $m_b$ and $N_c$=3, we obtain
a set of $a_1$, $a_4$, and $a_6$ as follows:
\begin{eqnarray}\label{set1}
a_1&=&1.05\;,
\nonumber\\
a_i&=&\big[(-a^0_i\mp8.5\eta-3.7\rho)+i(-115\pm3.7\eta-8.5\rho)\big]\times
10^{-4} \;, \ (i=4,6)
\end{eqnarray}
where $a^0_4=388$ and $a^0_6=556$ for $b\to s$ ($\bar b\to \bar s$) transition. As an illustration,
we would also like to turn off the strong phase ($\delta=0$), by
taking the imaginary parts of the quark-loop rescattering effects
to be zero. 

The numerical values for TPAs of $A_T$
($\bar A_T$) and ${\cal A}_T=(A_T-\bar{A}_T)/2$ are shown in
Table \ref{ATtable1}.
\begin{table}[h]
\tbl{Triple product correlation asymmetries (in percent)
of $A_T$ ($\bar A_T$) for $\bar{B}^0\to\Lambda\bar{p}\pi^+$
(${B}^0\to\bar{\Lambda} p\pi^-$) and ${\cal
A}_T=(A_T-\bar{A}_T)/2$.
}
{\begin{tabular}{@{}ccc@{}} \toprule
$A_T,\bar{A}_T,{\cal A}_T$&$\gamma =60^\circ$&$\gamma =0^\circ$\\
\colrule
$\delta\neq 0$&$9.9,\ -5.2,\ 7.6$&$3.4,\ 3.4,\ 0$\\
$\delta=0$&$10.4, -10.4, 10.4$&$0,\ \ 0,\ \ 0$\\
 \botrule
\end{tabular}}
\label{ATtable1}
\end{table}
From the table, we see explicitly that TPAs are indeed
nonzero and maximal in the absence of the strong phase. We note
that in our calculations we have neglected the final state
interactions due to electromagnetic and strong interactions, which are believed to be
small in three-body charmless baryonic decays \cite{GH05,ChengQCD}. It is
interesting to point out that in order to observe the TPAs in
$\bar B^0 \to \Lambda\bar p\pi^+$ and $B^0\to \bar\Lambda p \pi^-$
being at $10-7$\%, we need to have about $(1-2)\times 10^{8}$
$B\bar B$ pairs at $2\sigma$ level. This is within the reach of
the present day $B$ factories at KEK and SLAC and others that
would come up. It is clear that an experimental measurement of ${\cal A}_T$ is a reliable test of the CKM mechanism of CP violation and, moreover, it could be the first evidence of the direct T violation in B decays.
We remark that we  have also explored T violating effects in a large class of interesting charmless baryonic decays such as 
 $B\to\Lambda\bar{\Lambda}K$ \cite{GH05} and we have found that
 they are small \cite{GH-f}. 

Finally,
we  illustrate the difference between the TPAs and the rate CP asymmetry 
in Eq. (\ref{cpa}).
Nonzero contributions on $A_{CP}$ can be induced from the inteferences among
$M_1$, $M_4$ and $M_6$. 
We find that $A_{CP}$ for $\bar{B}^0\to\Lambda\bar{p}\pi^+$ is
$1.1\%$, which is about one order
of magnitude smaller than that of ${\cal A}_T$.
 It is clear that without strong or
weak phase there is no direct CP violating asymmetry as seen in Eq.
(\ref{cpa}).

\section*{Acknowledgements} This work is financially supported by
the National Science Council of Republic of China under the
contract  NSC-93-2112-M-007-014.

\end{document}